\documentclass[twocolumn]{aastex61}

\shorttitle{Transit Follow-Up of K2-22b}
\shortauthors{Col\'on et al.}

\usepackage{color}
\usepackage{graphics}
\usepackage{longtable}
 
\begin{document}

\title{A Large Ground-Based Observing Campaign of the Disintegrating Planet K2-22b}

\author{Knicole D. Col\'{o}n}
\affiliation{NASA Goddard Space Flight Center, Exoplanets and Stellar Astrophysics Laboratory (Code 667), Greenbelt, MD 20771, USA}

\author{George Zhou}
\affiliation{Harvard-Smithsonian Center for Astrophysics, Cambridge, MA 02138, USA}

\author{Avi Shporer}
\affiliation{Massachusetts Institute of Technology, Cambridge, MA 02139, USA}

\author{Karen A.\ Collins}
\affiliation{Harvard-Smithsonian Center for Astrophysics, Cambridge, MA 02138, USA}

\author{Allyson Bieryla}
\affiliation{Harvard-Smithsonian Center for Astrophysics, Cambridge, MA 02138, USA}

\author{N\'estor Espinoza}
\affiliation{Max-Planck-Institut f\"ur Astronomie, K\"onigstuhl 17, 69117 Heidelberg, Germany}
\affiliation{Instituto de Astrof\'isica, Facultad de F\'isica, Pontificia Universidad Cat\'olica de Chile, Av. Vicu\~na Mackenna 4860, 782-0436 Macul, Santiago,
Chile}
\affiliation{Millennium Institute of Astrophysics (MAS), Av. Vicu\~na Mackenna 4860, 782-0436 Macul, Santiago, Chile}

\author{Felipe Murgas}
\affiliation{Instituto de Astrof\'{i}sica de Canarias (IAC), E-38205 La Laguna, Tenerife, Spain}
\affiliation{Departamento de Astrof\'isica, Universidad de La Laguna (ULL), E-38205 La Laguna, Tenerife, Spain}

\author{Petchara Pattarakijwanich}
\affiliation{Department of Physics, Faculty of Science, Mahidol University, Bangkok 10400, Thailand}

\author{Supachai Awiphan}
\affiliation{National Astronomical Research Institute of Thailand, 260 Moo 4, Donkaew, Mae Rim, Chiang Mai, 50180, Thailand}


\author{James D. Armstrong}
\affiliation{Institute for Astronomy, University of Hawaii, 34 Ohia Ku St., Pukalani, Maui, HI, 96768, USA}

\author{Jeremy Bailey}
\affiliation{School of Physics, University of New South Wales, Sydney, NSW 2052, Australia}
\affiliation{Australian Centre for Astrobiology, University of New South Wales, Sydney, NSW 2052, Australia}

\author{Geert Barentsen}
\affiliation{NASA Ames Research Center, M/S 244-30, Moffett Field, CA 94035, USA}
\affiliation{Bay Area Environmental Research Institute, 625 2nd St. Ste 209, Petaluma, CA 94952, USA}

\author{Daniel Bayliss}
\affiliation{Department of Physics, University of Warwick, Coventry CV4 7AL, UK}

\author{Anurak Chakpor}
\affiliation{National Astronomical Research Institute of Thailand, 260 Moo 4, Donkaew, Mae Rim, Chiang Mai, 50180, Thailand}

\author{William D. Cochran}
\affiliation{McDonald Observatory and Department of Astronomy, The University of Texas at Austin,  Austin, TX 78712, USA}

\author{Vikram S. Dhillon}
\affiliation{Department of Physics and Astronomy, University of Sheffield, Sheffield, S3 7RH, United Kingdom}
\affiliation{Instituto de Astrof\'{i}sica de Canarias (IAC), E-38205 La Laguna, Tenerife, Spain}

\author{Keith Horne}
\affiliation{SUPA Physics and Astronomy, University of St Andrews, North Haugh, St Andrews KY16 9SS}

\author{Michael Ireland}
\affiliation{Research School of Astronomy \& Astrophysics, Australian National University, Canberra, ACT 2611, Australia}

\author{Lucyna Kedziora-Chudczer}
\affiliation{School of Physics, University of New South Wales, Sydney, NSW 2052, Australia}
\affiliation{Australian Centre for Astrobiology, University of New South Wales, Sydney, NSW 2052, Australia}

\author{John F. Kielkopf}
\affiliation{Department of Physics and Astronomy, University of Louisville, Louisville, KY 40292, USA}

\author{Siramas Komonjinda}
\affiliation{Department of Physics and Materials Science, Faculty of Science, Chiang Mai University, 239 Huay Kaew Road, Chiang Mai, 50200, Thailand}
\affiliation{Research Center in Physics and Astronomy, Faculty of Science, Chiang Mai University, 239 Huay Kaew Road, Chiang Mai, 50200, Thailand}

\author{David W.\ Latham}
\affiliation{Harvard-Smithsonian Center for Astrophysics, Cambridge, MA 02138, USA}

\author{Tom. R. Marsh}
\affiliation{Department of Physics, University of Warwick, Coventry CV4 7AL, UK}

\author{David E.\ Mkrtichian}
\affiliation{National Astronomical Research Institute of Thailand, 260 Moo 4, Donkaew, Mae Rim, Chiang Mai, 50180, Thailand}

\author{Enric Pall\'e}
\affiliation{Instituto de Astrof\'{i}sica de Canarias (IAC), E-38205 La Laguna, Tenerife, Spain}
\affiliation{Departamento de Astrof\'isica, Universidad de La Laguna (ULL), E-38205 La Laguna, Tenerife, Spain}

\author{David Ruffolo}
\affiliation{Department of Physics, Faculty of Science, Mahidol University, Bangkok 10400, Thailand}

\author{Ramotholo Sefako}
\affiliation{South African Astronomical Observatory, PO Box 9, Observatory, Cape Town 7935, South Africa}

\author{Chris G. Tinney}
\affiliation{School of Physics, University of New South Wales, Sydney, NSW 2052, Australia}
\affiliation{Australian Centre for Astrobiology, University of New South Wales, Sydney, NSW 2052, Australia}

\author{Suwicha Wannawichian}
\affiliation{Department of Physics and Materials Science, Faculty of Science, Chiang Mai University, 239 Huay Kaew Road, Chiang Mai, 50200, Thailand}

\author{Suraphong Yuma}
\affiliation{Department of Physics, Faculty of Science, Mahidol University, Bangkok 10400, Thailand}

\begin{abstract}

We present 45 ground-based photometric observations of the K2-22 system collected between December 2016 and May 2017, which we use to investigate the evolution of the transit of the disintegrating planet K2-22b. Last observed in early 2015, in these new observations we recover the transit at multiple epochs and measure a typical depth of $<$1.5\%. We find that the distribution of our measured transit depths is comparable to the range of depths measured in observations from 2014 and 2015. These new observations also support ongoing variability in the K2-22b transit shape and time, although the overall shallowness of the transit makes a detailed analysis of these transit parameters difficult. We find no strong evidence of wavelength-dependent transit depths for epochs where we have simultaneous coverage at multiple wavelengths, although our stacked Las Cumbres Observatory data collected over days-to-months timescales are suggestive of a deeper transit at blue wavelengths. We encourage continued high-precision photometric and spectroscopic monitoring of this system in order to further constrain the evolution timescale and to aid comparative studies with the other few known disintegrating planets. 

\end{abstract}


\section{Introduction}

Exoplanet surveys have yielded many surprises over the years. The discovery of ``disintegrating'' exoplanets was one such surprise. These are planets that appear to have tails of dusty material that produce asymmetric transit shapes. At present, there are only three such planets known around main-sequence stars: KIC 12557548b \citep{rappaport2012}, KOI-2700b \citep{rappaport2014}, and K2-22b \citep{Sanchis2015}. The first two were discovered in NASA's Kepler prime mission, while the latter was discovered in Campaign 1 of NASA's K2 mission \citep{howell2014}. Given that Kepler and K2 have observed a combined total of several hundred thousand stars, this suggests that such objects are either intrinsically rare or have a short enough survival lifetime that we are lucky to catch any in the act of disintegrating \citep{vanlieshout2017}.

Because such objects are rare, the systems named above have been under intense study so as to better understand their formation and evolution. In particular, observations over long timescales can be used to determine the rate at which the transit depth evolves over time.  In addition, multi-wavelength observations can provide constraints on the properties of the grains that are present in the dust tails. For example, several such studies have been done for WD 1145+017 \citep{vanderburg2015,vanderburg2018arXiv180401997V}, which is a white dwarf star that has disintegrating planetesimals in orbit around it and is perhaps the most well-studied ``disintegrating'' system to date \citep[e.g.,][]{alonso2016,gansicke2016,zhou2016,croll2017,hallakoun2017,redfield2017,vanderburg2018arXiv180401997V,xu2018}. However, because this system consists of debris orbiting a post-main-sequence star, it is arguably in a different class than the other three disintegrating planets known. 

The planetary companion KIC 12557548b, which orbits a highly-spotted K-dwarf  star with a period of $\sim$16 hr, displays variable transit depths ranging from $<$0.2\% to 1.3\% within Kepler data obtained between 2009 and 2013  \citep[e.g.,][]{rappaport2012}. It was later observed in 2013 and 2014 to have weaker transits overall than seen in the Kepler data \citep[e.g.,][]{schlawin2016}. Studies have found evidence for a correlation between the variability of the transit depth and the stellar rotation period \citep[$\sim$23 d;][]{kawahara2013,croll2015}. These studies suggest that either the activity corresponding to enhanced ultraviolet and/or X-ray radiation in turn causes increased mass-loss and therefore increased variability in the transit depth, or the apparent changes in transit depth occur as a dust tail passes over star spots. In addition, while simultaneous Kepler and near-infrared observations of KIC 12557548b revealed no significant difference in transit depth with wavelength \citep{croll2014}, evidence for a color dependence of the transit depth between $g^{\prime}$ and $z^{\prime}$ was later found by \citet{bochinski2015}. These observations provide an estimate on the grain size within the dust tail of 0.25$-$1 $\mu$m in radius.   

Similar to KIC 12557548b, the planetary object KOI-2700b orbits an active mid-K star with a period of $\sim$22 hr \citep{rappaport2014}. However, the transit depth is measured to be $<$0.04\% on average in the Kepler data, much shallower than measured for KIC 12557548b. Due to the shallowness of the transit, any transit-to-transit variability in the Kepler data is difficult to study.  Still, the transit depth was shown to be monotonically decreasing over the four-year timescale of the Kepler observations \citep{rappaport2014}. Unfortunately, further follow-up observations of this system are inherently difficult given the shallow transit. 

The planet K2-22b orbits an M-dwarf with a period of just $\sim$9 hr \citep{Sanchis2015}, much shorter than the two aforementioned disintegrating planets. The star itself exhibits photometric variability at the 1\% level with a rotation period of 15 days. \citet{Sanchis2015}  presented the analysis of the K2 observations that were obtained between May and August 2014 and 15 ground-based follow-up light curves obtained between January and March 2015. From these data, \citet{Sanchis2015} measured rapidly variable transit depths for K2-22b ranging from $\sim$0\% to 1.3\%, comparable to KIC 12557548b. A wavelength-dependent transit light curve shape was also measured during one particularly deep transit, supporting dust scattering during the transit. Another significant difference between K2-22b and both KIC 12557548b and KOI-2700b is that K2-22b appears to have both a leading and trailing dust tail, rather than just a trailing one. 

Given that the K2-22 system is unique among the few disintegrating planets known, we undertook a new observing campaign to further investigate the nature of this system. Here, we present 45 new ground-based light curves of the K2-22 system obtained between December 2016 and May 2017, 31 of which cover complete transit windows as predicted based on the ephemeris from \citet{Sanchis2015}. In Section \ref{obs} we describe these observations, which were obtained with telescopes ranging in size from 0.5 m to 10.4 m and spanning optical to near-infrared wavelengths. In Section \ref{analysis} we present our analysis and modeling of the light curves, and we describe and discuss our findings in Sections \ref{results} and \ref{discussion}. 

\section{Observations and Data Analysis}
\label{obs}

In the following sections we describe the observations of K2-22 performed with nine different facilities located around the world. The time-series photometry from each observatory is presented in Figure \ref{fig:lc_one} and Table \ref{tab:data}, and a summary of the observations is given in Table \ref{table:observations}.

\subsection{Anglo-Australian Telescope (AAT)}
Near-infrared light curves of K2-22 were obtained using the IRIS2 camera on the 3.9\,m Anglo-Australian Telescope \citep[AAT; ][]{Tinney2004}, located at Siding Spring Observatory (SSO) in Australia. IRIS2 is a $1\mathrm{K}\times1\mathrm{K}$ camera utilizing a HAWAII-1 HgCdTe detector, read out over four quadrants in the double-read mode, achieving a field of view of $7\farcm 7 \times 7\farcm 7$, at a pixel scale of $0\farcs 4486\,\mathrm{pixel}^{-1}$. 

Observations were obtained on UT 2017 March 15 (transit epoch = 2669) and UT 2017 March 16 (transit epoch = 2671) with IRIS2 in the $Ks$ band, at 30\,s exposure time. These observations were scheduled to accompany simultaneous optical photometry from the LCO 1\,m telescope at SSO. The IRIS2 observing procedure and photometric reductions largely follow that described in \citet{Zhou2014}. Dark exposures were subtracted from the data frames, and flat field division was performed with the aid of on-sky offset frames taken before and after the time series observations. To ensure the target star remained on the same pixel throughout the time series observations, we made use of an off-axis guider, as well as manual adjustments to the telescope pointing via real-time plate solutions from the science frames. Plate solving and final aperture photometric extraction were performed with the \textsc{fitsh} package \citep{Pal2012}. Relative photometry was performed with a set of four reference stars in the field. 

\subsection{Fred Lawrence Whipple Observatory (FLWO)}
We used KeplerCam on the 1.2 m telescope at the Fred Lawrence Whipple Observatory (FLWO) on Mt. Hopkins, Arizona to observe 5 transit windows of K2-22b. KeplerCam has a single 4K$\times$4K Fairchild CCD486 with a $0\farcs366$ pixel$^{-1}$ and a field of view of $23\farcm1 \times 23\farcm1$. The observations were made on UT 2017 March 8 (transit epoch = 2650), UT 2017 March 9 (transit epoch = 2652), UT 2017 April 16 (transit epoch = 2752), UT 2017 April 21 (transit epoch = 2765) and UT 2017 May 22 (transit epoch = 2846). All observations were obtained in an SDSS $i$ filter with a 120 s exposure time. 
Images were reduced using standard \textsc{idl} routines as outlined in \citet{carter:2011}. \textsc{AstroImageJ} \citep{collins2017} was used to perform aperture photometry on the processed images. We used an aperture radius of 8 pixels, corresponding to a $2\farcs93$ radius.

\subsection{Gao Mei Gu Observatory (GAO)}

We observed 3 transits of K2-22 on 3 consecutive nights from UT 2017 March 15 to UT 2017 March 17 (transit epochs = 2669, 2672, 2674) using the 0.7 m Thai Robotic Telescope - Gao Mei Gu Observatory (TRT-GAO), in Lijiang, China, with an Andor iKon-L 936 2K $\times$ 2K CCD camera with a scale of $0\farcs613$ pixel$^{-1}$. The observations were performed through a Cousins-I filter with 60 s exposures. The total science images on each night are 250 (UT 2017 March 15), 140 (UT 2017 March 16) and 150 (UT 2017 March 17). The calibration was carried out using the \textsc{iraf} tasks along with astrometric calibration using Astrometry.net \citep{lang2010}. The aperture photometry was carried out using \textsc{sextractor} \citep{bertin1996} with an adaptive scaled aperture based on the seeing in an individual image. The final apertures used were between 4-5 pixels equivalent to $2\farcs45$$-$$3\farcs07$ \citep{bertin1996}.

\subsection{Gran Telescopio Canarias (GTC)}
We were awarded Director's Discretionary Time to use the OSIRIS instrument \citep{Cepa2000} mounted on the 10.4 m Gran Telescopio Canarias (GTC) to observe 1 transit of K2-22b (PI: E. Pall\'e). The data were obtained on UT 2017 May 17 (transit epoch = 2835) using OSIRIS long-slit spectroscopy mode with the R1000R grism (spectral coverage of 510 nm - 1000 nm), $2\times 2$ binning mode ($0\farcs254$ pixel$^{-1}$), readout speed of 200 kHz, gain of 0.95 e$^{-}$/ADU, and readout noise of 4.5 e$^{-}$. A custom built long slit with a width of 12$^{\prime\prime}$ was used to obtain spectra of K2-22 and one reference star. The exposure time was set to 250 s and the total observing time was 3.3 hr (airmass varied from 1.12 to 1.97). A total of 45 science images was acquired during the run. Images were reduced using standard procedures (bias and flat calibration), and the spectral extraction of K2-22 and the reference star was done using an aperture of 40 binned pixels ($10\farcs2$) in width. For more details on the GTC data reduction procedure we refer the reader to Section 7 of \citet{Sanchis2015}. 

\subsection{Las Cumbres Observatory (LCO)}

We collected 24 light curves of K2-22 between UT 2017 March 1 and UT 2017 May 29 in either the Sloan
$g$ or Sloan $i$ band from multiple 1 m telescopes in the Las
Cumbres Observatory (LCO) network \citep{brown2013}. The corresponding transit epochs and site of each observation are given in Table \ref{table:observations}, where CTIO refers to Cerro Tololo Inter-American Observatory in Chile, McDonald refers to McDonald Observatory in Texas, SAAO refers to South African Astronomical Observatory in Sutherland, South Africa, and SSO refers to Siding Spring Observatory near Coonabarabran, New South Wales, Australia. The 1 m LCO telescopes each have a 4 K $\times$ 4 K Sinistro detector with a 26$\arcmin$ $\times$ 26$\arcmin$ field of view and a pixel scale of $0\farcs39$ pixel$^{-1}$. Calibrated data were downloaded from the LCO archive and then analyzed to extract the photometry following a similar procedure as described in the following section.

\subsection{The Swope Telescope}
We collected two light curves of K2-22 on UT 2017 March 13 (transit epoch = 2663) and UT 2017 April 11 (transit epoch = 2741) using the E2V 4K$\times$4K CCD at the 40-inch Swope telescope at Las Campanas Observatory. Observations were acquired in both cases with the SDSS $i$ filter but with different exposure times. Due to the closeness of the target to the Moon on the March 13 observations, the exposure times were set at 70 seconds. This allowed us to achieve photometry at the $\sim$2\% level only, due to the local sky brightness. The observations spanned almost 5 hours, covering a significant portion of the orbit of K2-22b.  On April 11 the Moon did not interfere with our observations and thus the exposure times were set to 110 seconds, which allowed us to reach photometry at the 0.5\% level per point over 6.5 hours, covering a significant fraction of the orbit of K2-22b. The data were reduced using a standard photometric pipeline which makes use mainly of tools from the Astropy package \citep{astropy2013}, and performs bias, dark and flat field corrections, along with astrometric identification of stars in the field using Astrometry.net \citep{lang2010} and subsequent extraction of photometry for all the stars in the field. Differential photometry for K2-22 was produced by using an ensemble of 10 comparison stars over a 10 pixel radius, which implies a radius of $4\farcs35$ as the pixel scale is $0\farcs435$ pixel$^{-1}$.

\subsection{Thai National Observatory (TNO)}

At Thai National Observatory (TNO), in Chiang Mai, Thailand, we conducted photometric observations of K2-22b on the 0.5 m Thai Robotic Telescope - Thai National Observatory (TRT-TNO) and the 2.4 m Thai National Telescope (TNT) on UT 2017 March 15 (transit epoch = 2669) and UT 2017 April 13 (transit epoch = 2745), respectively. 

A transit observation was conducted with the TRT-TNO through a Cousins-R filter using an Andor iKon-L 936 2K $\times$ 2K CCD camera attached to the 0.5 m Schmidt-Cassegrain Telescope. The field-of-view of each image is 23.4$^{\prime}$ $\times$ 23.4$^{\prime}$. The observation was conducted using 60 s exposure time and 250 science images were obtained. The calibration and photometry were performed using the same routine as TRT-GAO observations.

For the TNT observations, the target was observed with ULTRASPEC (Dhillon et al. 2014) in the $i$-band. ULTRASPEC is a high-speed camera using a frame-transfer electron-multiplying CCD (EMCCD). The target was observed in full-frame mode, with a field-of-view of 7.7$^{\prime}$ $\times$ 7.7$^{\prime}$ and a pixel scale of $0\farcs45$ pixel$^{-1}$. The exposure time was 42.78 s, with a dead time of 15 ms between each frame. More than 400 science images were obtained, but only the first 221 images are used in our analysis as latter images were affected by clouds. Plate matching and aperture photometric extraction were performed using the \textsc{fitsh} package \citep{Pal2012}. Relative photometry utilized a largely similar set of reference stars as the LCO observations (barring differences due to the different field of view). Light curves were extracted over six different apertures and background annulus radii to optimize the point-to-point photometric scatter.

\subsection{University of Louisville Manner Telescope (ULMT)}

We observed five complete transit windows of K2-22b on UT 2017 March 08 (transit epoch = 2650), UT 2017 March 09 (transit epoch = 2652), UT 2017 March 15 (transit epoch = 2668), UT 2017 May 22 (transit epoch = 2846), and UT 2017 May 25 (transit epoch = 2854) using the University of Louisville Manner Telescope (ULMT) located at the Mt. Lemmon summit of Steward Observatory, AZ. The observations employed a 0.6 m f/8 RC Optical Systems Ritchey-Chr\'{e}tien telescope and an SBIG STX-16803 CCD with a $4{\rm K}\times4{\rm K}$ array of 9\,$\mu$m pixels, yielding a $26\farcm6 \times 26\farcm6$ field of view and $0\farcs39$ pixel$^{-1}$ image scale. All observations were conducted using 200\,s exposure times and no filter. Conditions were clear except for some thin clouds during the observations on UT 2017 March 09 and UT 2017 March 15, which caused higher scatter in the light curves.

We used \textsc{AstroImageJ} \citep{collins2017} to perform bias, dark, and flat-field corrections, and to perform fixed radius circular aperture photometry on the final processed images. We used aperture radii of 10 pixels ($3\farcs9$) and a comparison ensemble of four or more stars to produce differential target star light curves.

\subsection{Wisconsin-Indiana-Yale-NOAO Observatory (WIYN)}
We observed one complete transit window of K2-22b on UT 2016 December 13 (transit epoch = 2427) using the WIYN\footnote{The WIYN Observatory is a joint facility of the University of Wisconsin-Madison, Indiana University, the National Optical Astronomy Observatory and the University of Missouri.} High-Resolution Infrared Camera (WHIRC) installed on the 3.5 m WIYN telescope at Kitt Peak National Observatory in Arizona. WHIRC is a Raytheon Virgo HgCdTe detector with a 2048 $\times$ 2048 array, a pixel scale of $0\farcs1$ pixel$^{-1}$, and a $3\farcm4 \times 3\farcm3$ field of view.  The $Ks$-band observations began about one hour before the expected transit and ended approximately one hour after the end of the expected transit, when morning twilight began. Some clouds were present during the observations. 

The first step in processing the WHIRC data involved using the \emph{wprep}  \textsc{iraf} script available on the WHIRC instrument website\footnote{https://www.noao.edu/kpno/manuals/whirc/WHIRC.html} to trim the images and perform a linearity correction. We then corrected the images against darks, masked bad pixels, removed the pupil ghost using an image generated by dividing $Ks$ and $J$ flats, and performed flat-fielding. We used \textsc{AstroImageJ} \citep{collins2017} to perform circular aperture photometry on the final processed images, using an aperture radius of 13 pixels ($1\farcs3$) and one comparison star to produce a light curve.

\begin{figure*}[h!]
\begin{center}
\includegraphics[scale=0.86,angle=0]{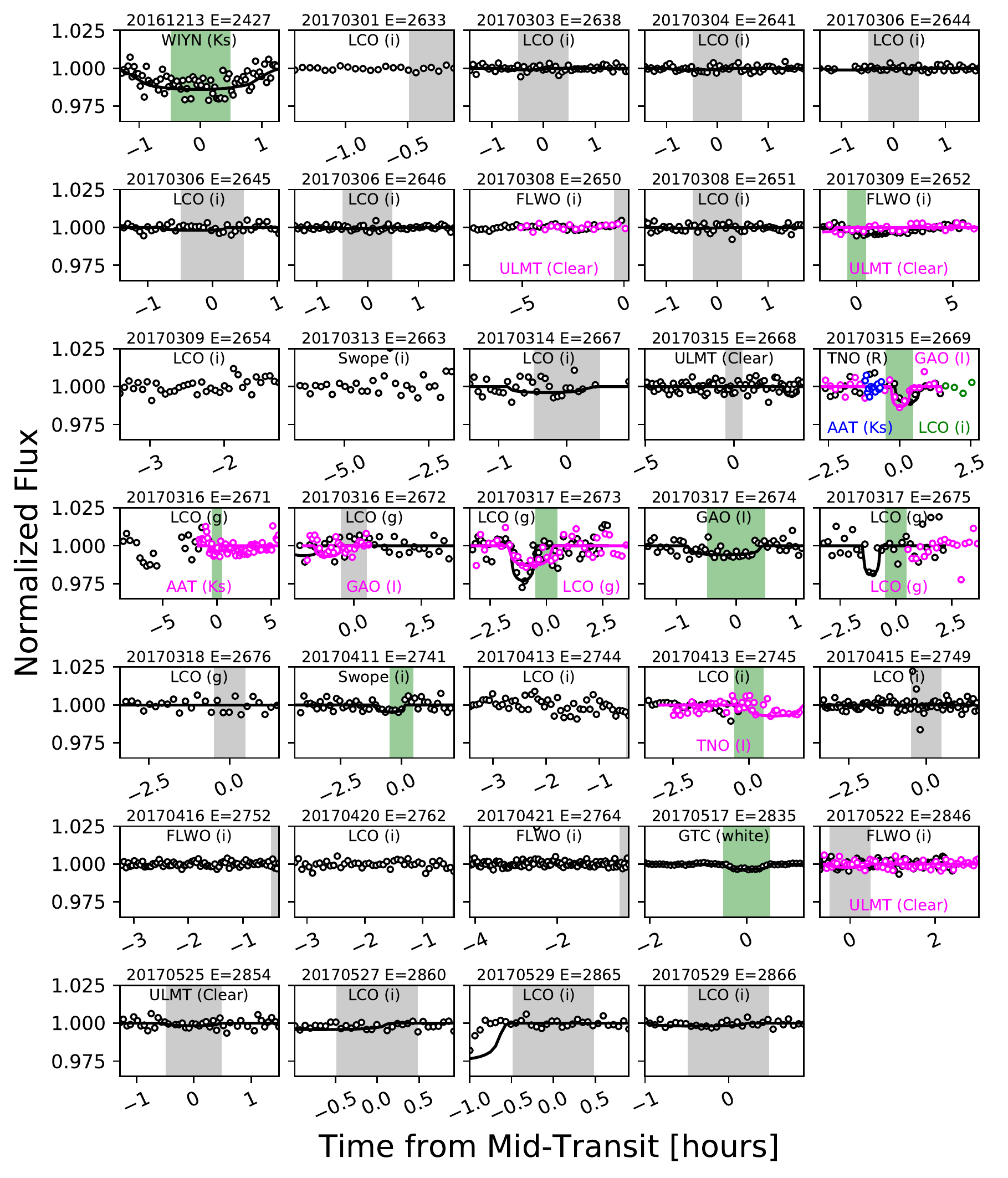}
\caption{Ground-based photometric follow-up light curves of K2-22 (open circles). Each panel lists the facility and filter used to collect the data as well as the start date of the observations (UT) and the transit epoch, relative to the transit ephemeris from \citet{Sanchis2015}. To aid visual comparisons between different light curves, all panels have the same vertical axis, and some light curves have been binned for clarity. As shown in some panels, data from different observatories covered the same epoch and are plotted in different colors. The green or gray shaded regions in some panels mark the transit windows of K2-22b based on the ephemeris and duration (48 min) from \citet{Sanchis2015}. The panels with shaded green regions specifically highlight epochs where we were able to robustly measure a transit depth (see Table \ref{table:observations}). Solid lines are the models that were fitted to the light curves that spanned a complete transit window (see \S\ref{analysis} for details). The data in the second to last panel on the bottom show a significant trend at the beginning of the observations, which is likely not astrophysical and adversely affected the best-fitting model (see \S\ref{results} for details).      \label{fig:lc_one}}
\end{center}
\end{figure*}






\begin{deluxetable}{ccccc}
\tablecaption{Time-Series Photometry of K2-22 \label{tab:data}}
\tablehead{
\colhead{Observatory} & \colhead{Filter} & \colhead{BJD} & \colhead{Relative} & \colhead{Relative Flux}\\
&  & & \colhead{Flux} & \colhead{Error}
}
\startdata
WIYN & Ks & 2457735.942395 & 0.999131 & 0.009729 \\
WIYN & Ks & 2457735.942858 & 0.998627 & 0.009729 \\
WIYN & Ks & 2457735.943321 & 0.991867 & 0.009729 \\
WIYN & Ks & 2457735.943773 & 1.004635 & 0.009729 \\ 
WIYN & Ks & 2457735.944236 & 0.993868 & 0.009729 \\
... & ... & ... & ... & ... \\
\enddata
\tablecomments{A portion of the table is shown here.  The full version containing photometry from all observatories is available online.}
\end{deluxetable}  

\section{Analysis}
\label{analysis}

\subsection{Light Curve Modeling}

To better understand the distribution of transit depths in our data and our detection limits, we performed a transit light curve fit using the \citet{mandel2002} model to all 31 data sets that covered most of the expected transit window \citep[as per the ephemeris from][]{Sanchis2015}. The following constraints were imposed on the fits: the transit duration\footnote{The transit duration previously measured from K2 and ground-based data is 48$\pm$3 min \citep{Sanchis2015}.} was required to be $<$2 hr, the planet-star radius ratio ($R_p/R_*$) to be $<$1, and the mid-transit time should be within 1 hr of the predicted transit time. To account for long-term trends in the baseline of a light curve, we also allowed for a quadratic trend to be fitted simultaneous to the transit fit. The resulting values measured for $R_p/R_*$ are given in Table \ref{table:observations}. If no uncertainty in $R_p/R_*$ is provided in the table, then the value for $R_p/R_*$ is the 1$\sigma$ upper limit.  If no value for $R_p/R_*$ is given at all, then that particular light curve did not cover a complete transit window of K2-22b. Figure \ref{fig:lc_one} shows the detrended data and best-fitting models to the 31 light curves that fit our criteria. 

\subsection{Impact of the Nearby Neighbor}

As identified in \citet{Sanchis2015}, K2-22 has a faint nearby companion located $\sim$2$^{\prime\prime}$ away. The angular proximity of this neighboring star means that in most of our photometry, the flux from the target and the neighbor are blended. To mitigate the effects of the neighboring star, we follow \citet{ciardi2015} to calculate the true planet radius (and therefore the true planet-star radius ratio): $R_p$(true)/$R_p$(observed) = $\sqrt[]{F_{total}/F_1}$, where $F_1$ is the flux of K2-22 as it is the star that is being transited and $F_{total}$ is the combined flux of the primary and neighbor star. We used magnitudes provided in Table 3 of \citet{Sanchis2015} for K2-22 and its neighbor to compute $F_1$ and $F_{total}$ for the relevant bandpasses where we measured either a detection or upper limit of the transit of K2-22b. We provide the corrected planet-star radius ratios in Table \ref{table:observations}. 

Given the proximity of the neighbor, there is some probability that it is the source of the transit signal rather than K2-22. However, \citet{Sanchis2015} argue why K2-22 is the most likely transit host rather than the neighbor. Furthermore, the WIYN observations presented here resolved K2-22 from the neighbor and provided a tentative detection of the K2-22b transit. More importantly, the neighbor did not display an obvious deep transit during the WIYN observations, which is additional evidence it is not the source of the transit signal.

\section{Results}
\label{results}

\subsection{Long-Term Monitoring of the Transit Depth}

\begin{figure*}[h!]
\begin{center}
\includegraphics[scale=0.7,angle=0]{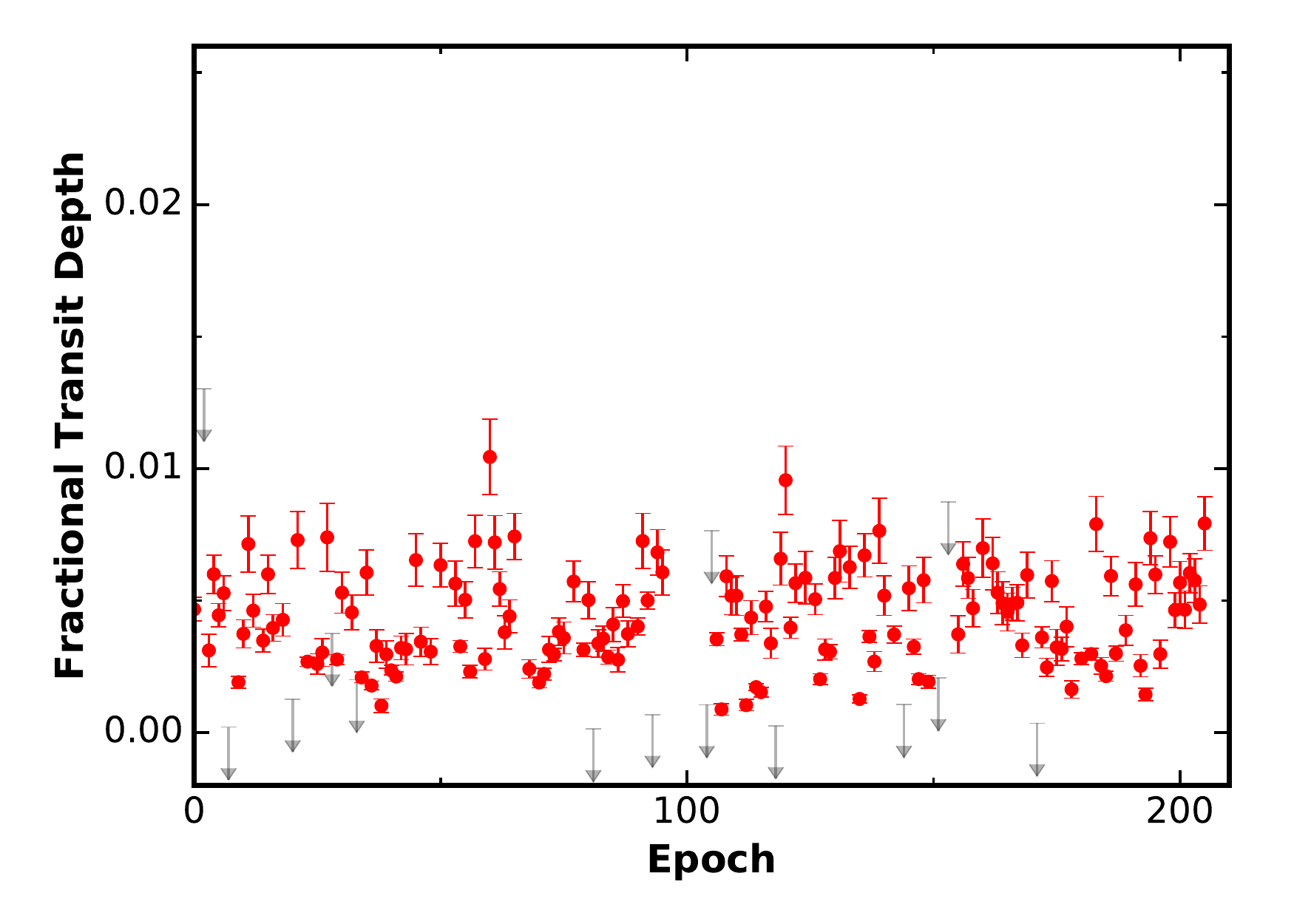}
\caption{Fractional transit depth as a function of epoch for K2-22b measured from K2 Campaign 1 data collected between May and August 2014. The epoch is relative to the ephemeris and orbital period from \citet{Sanchis2015}: 2456811.1208 BJD and 9.145872 hr. Red points and error bars are detections of the transit, while gray markers with downward facing arrows are upper limits. The vertical axis is the same scale as in Figure \ref{fig:depths} to aid visual comparison between the measured K2 and follow-up transit depths. \label{fig:depths_k2}}
\end{center}
\end{figure*}

To determine how our measured transit depths compare to previous observations, we returned to the K2 Campaign 1 discovery data collected between May and August 2014 and followed the procedure outlined in \S\ref{analysis} to measure the depths of transits in that long-cadence data (where each measurement has an integration time of 30 min). Depths were measured for transits that had at least 2 points in-transit, and the results are shown in Figure \ref{fig:depths_k2} as a function of epoch. The detected K2 transits have depths ranging from 0.21$-$0.57\% (the 1$\sigma$ distribution) with a median depth of 0.37\%. 

\begin{figure*}[h!]
\begin{center}
\includegraphics[scale=0.7,angle=0]{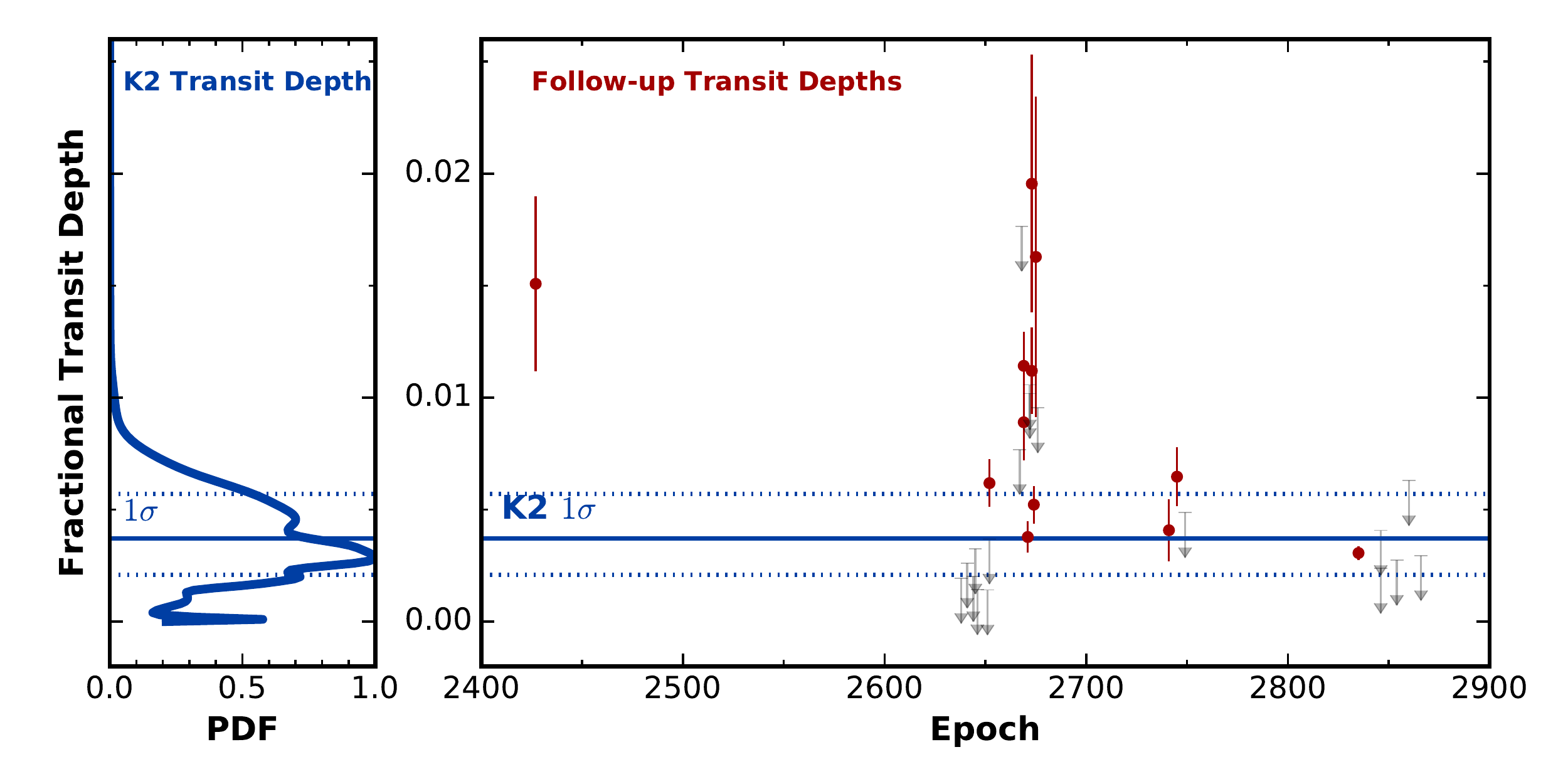}
\caption{{\bf Left:} Probability Distribution Function of the transit depths measured from K2 Campaign 1 data (Figure \ref{fig:depths_k2}). The horizontal solid line marks the median transit depth measured from K2, while the horizontal dotted lines indicate the 1$\sigma$ distribution of the measured depths. {\bf Right:} Measured transit depth from our follow-up photometry of K2-22b as a function of epoch. As in Figure \ref{fig:depths_k2}, the epoch is relative to the ephemeris and orbital period from \citet{Sanchis2015} and red points and error bars are detections of the transit, while gray markers with downward facing arrows are upper limits. The transit depths have been corrected for the flux contribution from the companion star. The transit depths we measure in our follow-up data are consistent with those measured from K2 data, within the uncertainties.  \label{fig:depths}}
\end{center}
\end{figure*}

In Figure \ref{fig:depths}, we show the measured transit depth (or upper limit on the transit depth) as a function of epoch from our ground-based follow-up campaign, along with the Probability Distribution Function (PDF) of the transit depths measured from K2 data. Due to the overall shallowness of the transit, we detected the transit in just $\sim$one-third of the 31 complete transit windows observed. We still find that our measured depths are consistent with what K2 detected, within the uncertainties. 

\begin{figure*}[h!]
\begin{center}
\includegraphics[scale=0.8,angle=0]{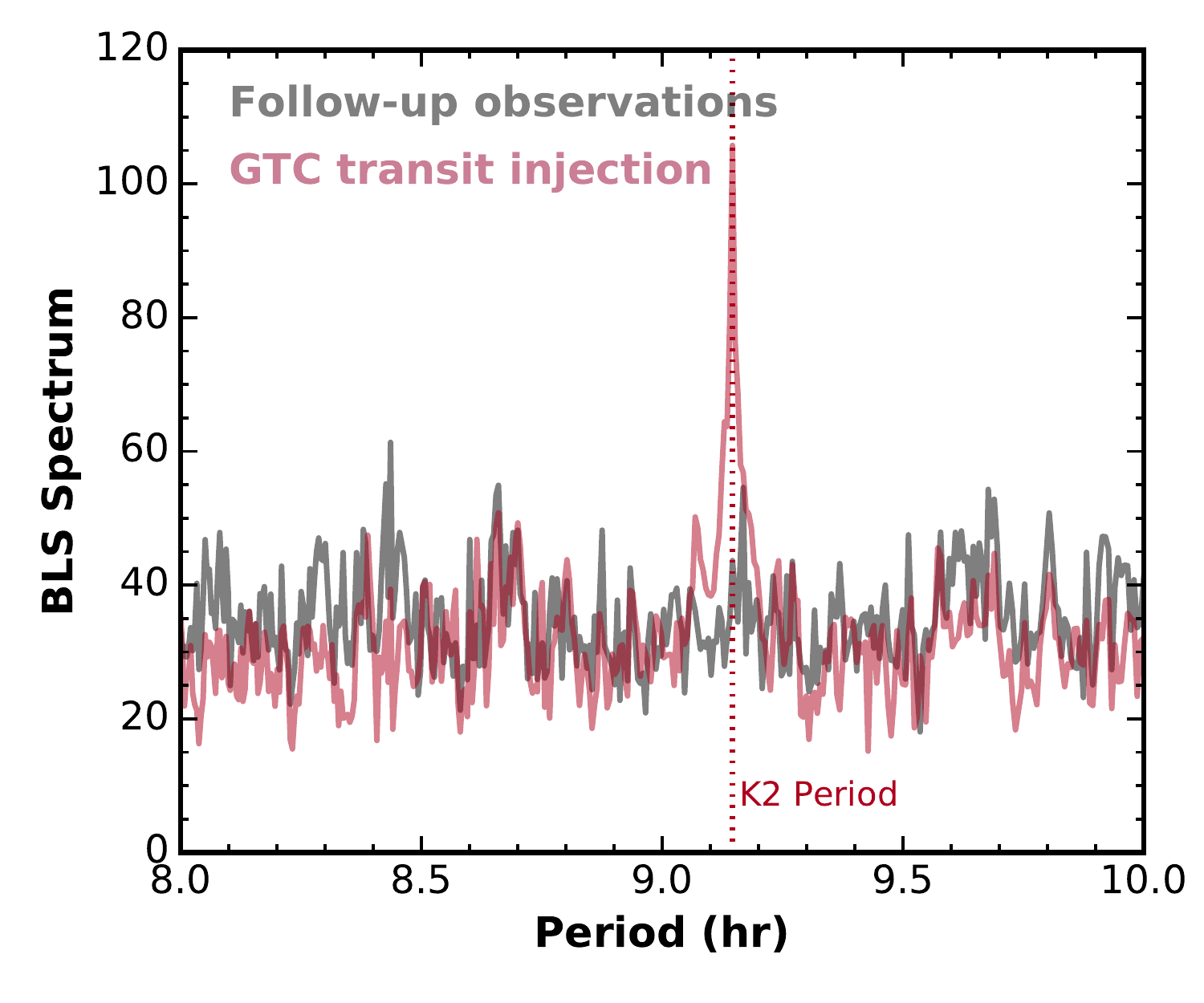}
\caption{BLS spectrum of the light curves presented in Figure \ref{fig:lc_one}. The gray curve shows that no clear periodicity is seen in our data at the period of K2-22b (marked with the vertical dashed line). The red curve shows that a periodic signal is recovered at the period of K2-22b when injecting a transit with the same depth as the GTC-measured transit into our other data. 
\label{fig:bls}}
\end{center}
\end{figure*}

Our most robust detection was obtained from the GTC, with a measured transit depth of $\sim$0.3\% (Figure \ref{fig:lc_one}). To test whether we should have recovered a transit of this depth in our other individual data sets, we injected the GTC signal into the other light curves we collected. The resulting Box-fitting Least Squares \citep[BLS;][]{kovacs2002} spectrum is shown in Figure \ref{fig:bls}. When performing the BLS search on our data, we searched for transits with $q_{min}$ = 0.02, $q_{max}$ = 0.15, and orbital periods between 0-10 days, where $q$ is the transit duration/orbital period. Figure \ref{fig:bls} illustrates that if the transit of K2-22b had a consistent depth of $\sim$0.3\%, we would have recovered it in the other individual observations. This suggests that the transit depth of K2-22b was changing over the course of our campaign. Figure \ref{fig:depths} further supports this: for example, an initial detection of the K2-22b transit was made in December 2016 followed by a cluster of non-detections (upper limits) in early March 2017 (epoch 2638), and then beginning in mid-March 2017 (epoch 2669) we had several positive detections of the transit. Several of the light curves where we had a non-detection had sufficiently high photometric precision that we should have recovered a $\sim$0.3\% transit. Therefore, we conclude that over the course of our observing campaign, the transit depth was indeed changing and was at times deeper and at other times shallower than the highest-precision event obtained by the GTC on 2017 May 17 (Figure \ref{fig:lc_one}). 

\begin{figure*}[h!]
\begin{center}
\includegraphics[scale=0.6,angle=0]{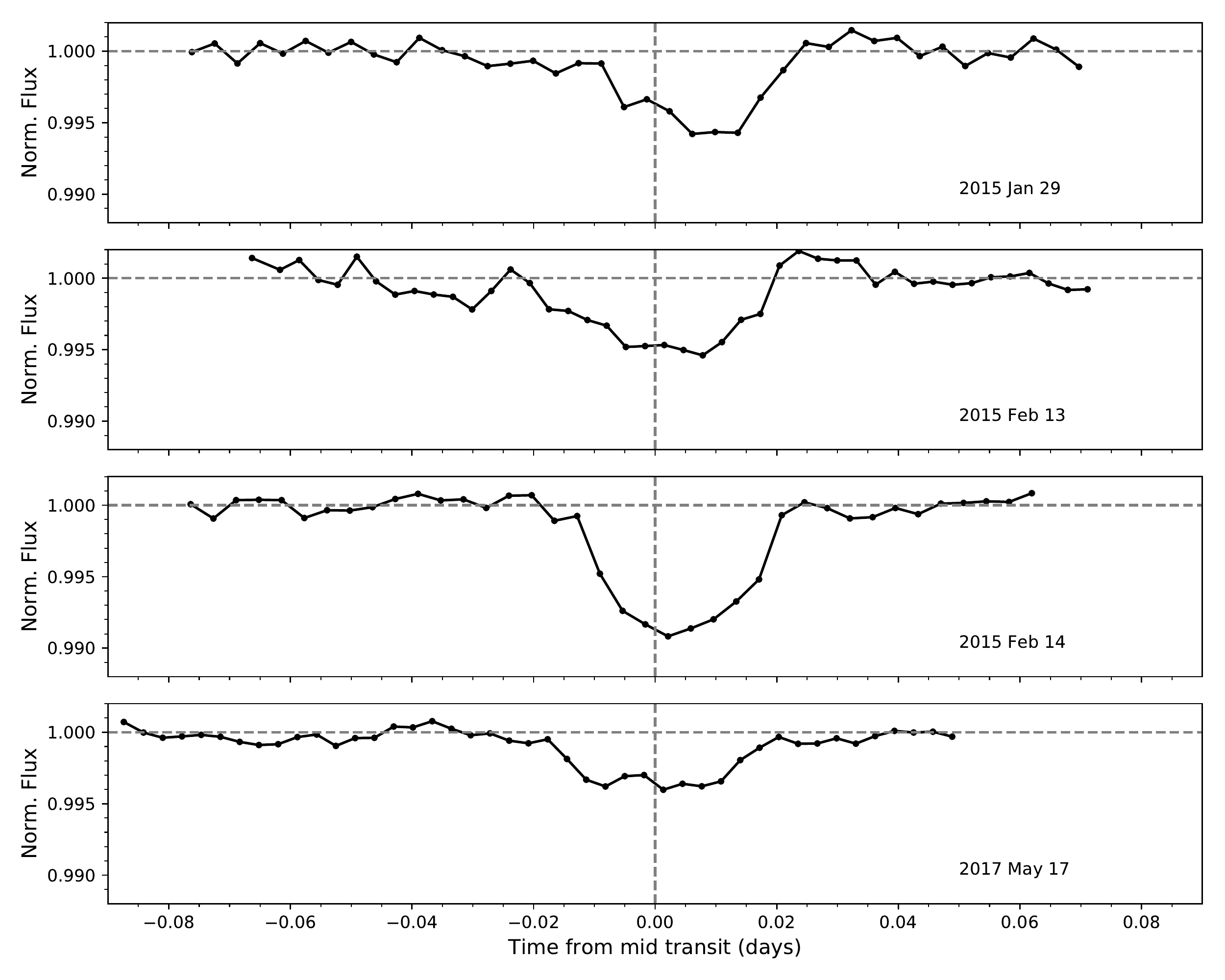}
\caption{White light curves of K2-22b from the GTC. The top three panels show data from 2015 (transit epoch =  634, 673, 676, respectively), originally presented in \citet{Sanchis2015}. The bottom panel shows the data from 2017 (transit epoch = 2835), and it is clear that this transit is shallower than previously recorded. However, this latest transit remarkably takes place right at the predicted time, and it also appears to be more symmetric than previous transits. \label{fig:gtc_data}}
\end{center}
\end{figure*}

Comparing the high-precision transits observed with the GTC in 2015 and in 2017 (Figure \ref{fig:gtc_data}) also reveals that the 2017 transit is shallower than previously measured in 2015, however, the 2017 observations are still consistent with depths measured from K2 overall. This instead suggests that the 2015 transits may have been observed at a time of relatively higher activity (e.g., enhanced ultraviolet and/or X-ray radiation) that led to increased mass-loss and deeper transits than typically seen by K2. 

One additional note of interest is that the GTC transit observed in 2017 appears to have occurred exactly when predicted and to be more symmetric than the previously observed transits from the GTC. The other light curves presented in Figure \ref{fig:lc_one} have poorer individual photometric precision than the GTC, so we are unable to perform a robust investigation of any changes in the transit shape or ephemeris over each epoch.  While some appear to have longer durations than expected and/or occur outside the predicted transit window, due to the relatively poor photometric precision in those cases we do not claim that those deviations in transit duration or time are significant. 

\begin{figure*}[h!]
\begin{center}
\includegraphics[scale=0.7,angle=0]{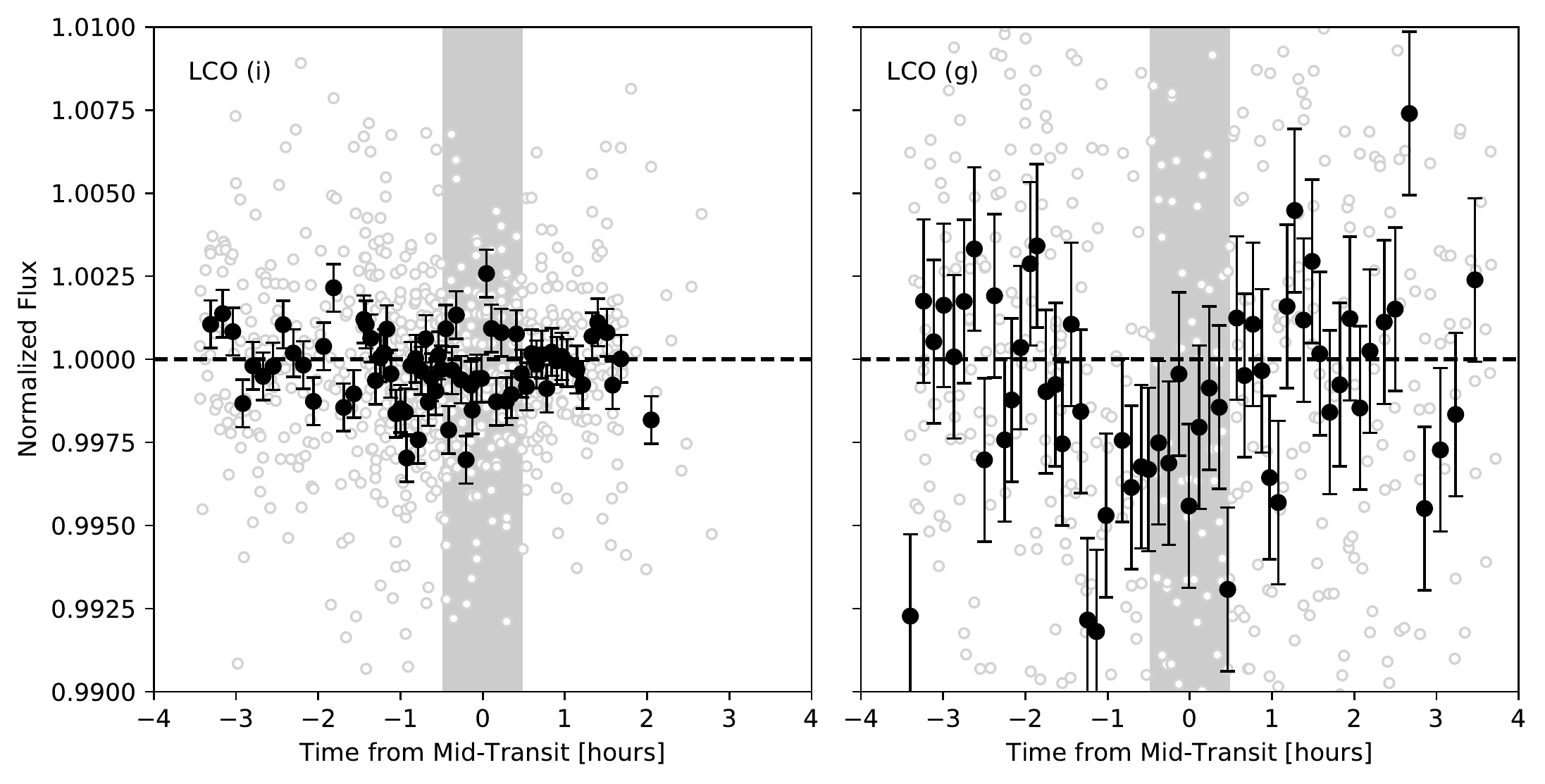}
\caption{Stacked and phased light curves from LCO in the Sloan $i$ band \textbf{(left)} and $g$ band \textbf{(right)}. In total, 17 $i$ band light curves were stacked in the left panel and 7 $g$ band light curves were stacked in the right panel. The full data sets are shown in light gray open circles, while binned light curves are shown as blacked filled circles. The dark gray shaded region indicates the predicted time and duration of the transit of K2-22b, and the horizontal dashed line marks where the normalized flux = 1 to guide the eye. All $g$ band data were obtained between UT 2017 March 16-18 over five transit epochs (2671$-$2676), while the $i$ band data were collected between UT 2017 March 1 and May 29. The LCO $i$ band data do not show an obvious transit. While the LCO $g$ band data display evidence of a $\sim$0.3\% transit that occurs slightly earlier than predicted, it has significantly larger photometric error bars than the $i$ band data.  \label{fig:lco}}
\end{center}
\end{figure*}

Because we have 24 light curves from LCO in total (17 in Sloan $i$ band and 7 in Sloan $g$ band), we explored whether stacking the phased light curves would provide a clear detection of the K2-22b transit. Figure \ref{fig:lco} shows the stacked light curves from LCO in both bands, where the photometric error bars are defined as the standard deviation of the binned light curve outside of the expected transit window. Interestingly, while we do not see an obvious transit in the $i$ band data, the $g$ band data reveal a potential $\sim$0.3\% transit occurring slightly earlier than predicted. We note the $g$ band data were collected between UT 2017 March 16-18 between transit epochs 2671 and 2676, while the $i$ band data were collected over several months between UT 2017 March 1 and May 29. Since the $g$ band data were collected over a relatively short timescale, this suggests the $g$ band detection is robust, although these data have notably poorer photometric precision than the $i$ band data. The $g$ band depth is also consistent with what we measured from the high-precision GTC light curve collected two months later on UT 2017 May 17. However, given the significant scatter in the $g$ band data, we cannot definitively conclude that an early transit of K2-22b was detected in the stacked light curves.  

In addition, there are a few individual measurements that may be anomalous and therefore affecting our interpretation of the data. The first detection in December 2016 (with WIYN; see Figure \ref{fig:lc_one}) is highly dependent on detrending. If we assume the transit is longer than expected, then the curvature seen in the light curve could be indicative of a transit as we present here. On the other hand, if we assume the transit has the expected duration of $\sim$48 min, then detrending the data results in a null detection of a transit and reduces the apparent slope in the transit depth over time. The uncertainty in the transit duration and depth for the WIYN data likely stems from the fact that we only had one suitable comparison star in the field of view. LCO observations at epoch 2865 (Figure \ref{fig:lc_one}) also show a significant trend at the beginning of the light curve, but as the data acquired at the subsequent epoch do not show any anomalous trend, we believe this is not astrophysical and instead some type of instrumental artifact. The best-fitting model is correspondingly a poor fit given this anomalous light curve feature.

\subsection{Search for Wavelength-Dependent Transit Light Curve Shape}

The few epochs where we have multi-wavelength coverage over a complete transit window (i.e., epochs 2652, 2669, 2672, 2846 in Figure \ref{fig:lc_one}) do not reveal any significant color dependence. While the $R$ and $I$ band data at epoch 2669 are suggestive of a depth difference, we do not have the photometric precision to make a robust claim. 

Figure \ref{fig:lco} shows the stacked light curves from LCO in Sloan $i$ and $g$ bands, and there we see that the $g$ band data are suggestive of a $\sim$0.3\% transit. On the other hand, we see no clear evidence for a transit in the $i$ band data. That we find evidence of a chromatic transit that is deeper at bluer wavelengths is consistent with what has been observed for, e.g., KIC 12557548b \citep{bochinski2015} and KIC 8462852 \citep{boyajian2018,deeg2018} (see Section \ref{discussion} for further discussion of these other systems). The caveat is that we are comparing stacked light curves collected over several days to months, and the $g$ band data have poorer photometric precision than the $i$ band data. We therefore cannot make any definitive claims about the chromaticity based on the LCO data and instead conclude that the LCO data are \emph{suggestive} of a deeper transit at blue wavelengths.

\begin{figure*}[h!]
\begin{center}
\includegraphics[scale=0.6,angle=0]{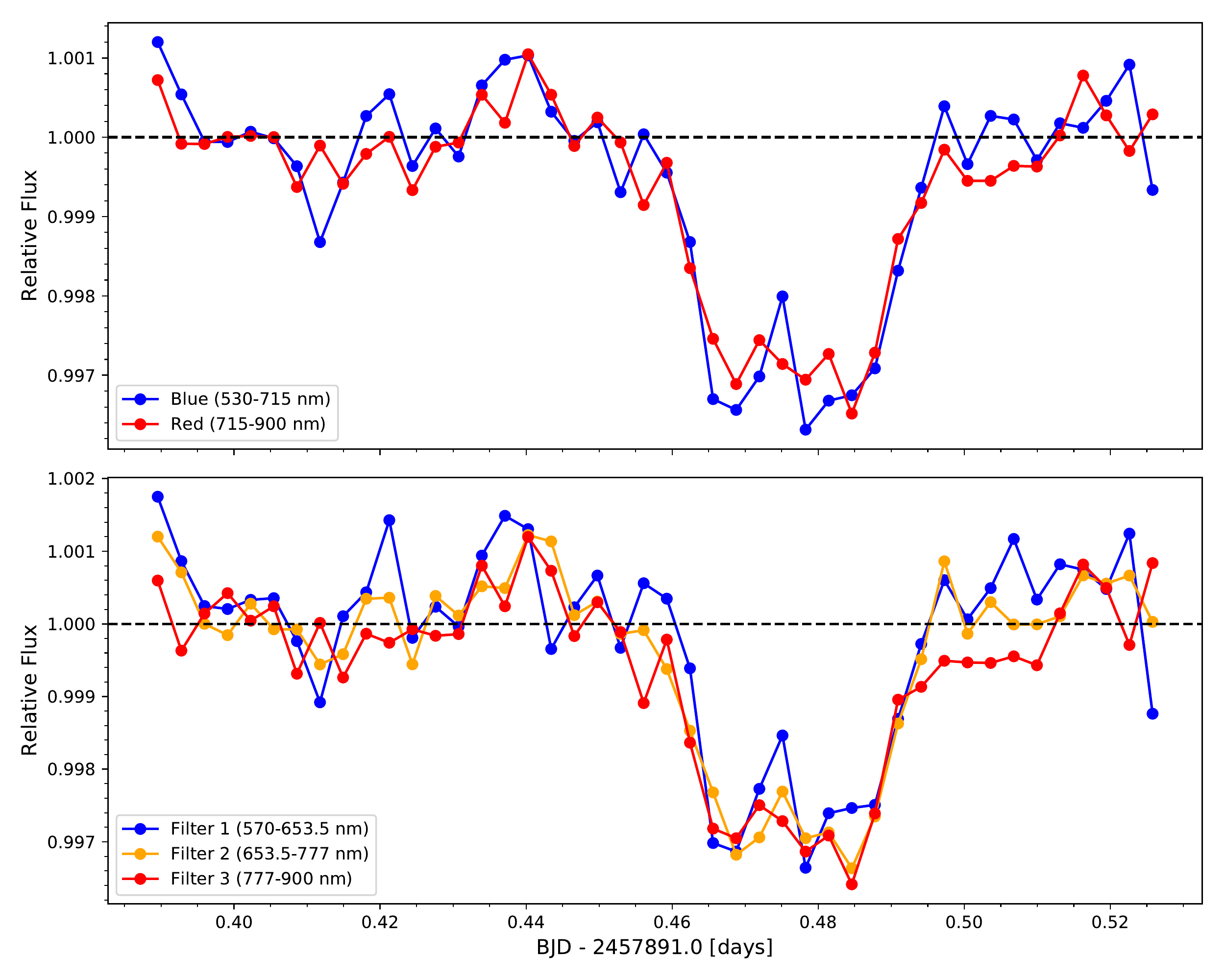}
\caption{GTC observations of K2-22 from UT 2017 May 17 (transit epoch = 2835) split among blue and red wavelengths (top panel) and among three color bands (bottom panel). No significant variations in the color of the transit are detected in either case. \label{fig:gtc_color}}
\end{center}
\end{figure*}

We also analyzed the spectroscopic data collected from the GTC, where in Figure \ref{fig:gtc_color} we present the GTC data split into blue and red wavelengths and also in three different color bands. During this one transit, no significant color dependence in the depth is measured. We note that of the three transits observed previously with the GTC, only one displayed a color dependent depth (with a deeper transit seen at blue wavelengths than red), and that was the deepest transit measured on UT 2015 Feb 4 as shown in Figure \ref{fig:gtc_data} and \citet{Sanchis2015}. \citet{ridden2018} recently demonstrated that the transit depth of disintegrating rocky planets is wavelength-independent for optically thick tails. The single wavelength-dependent transit observed for K2-22b to date could therefore be a sign of an optically thin tail at that particular epoch.  

\section{Discussion}
\label{discussion}

We have presented 45 new ground-based light curves of the K2-22 system collected in December 2016 and between March and May 2017. These light curves span 34 individual transit epochs of K2-22, and we were able to measure a transit depth for 12 of these epochs. We find that the measured transit depths of K2-22b over this period are variable at a level that is consistent with the range of transit depths measured by the K2 mission between May and August 2014 as well as with the ground-based campaign conducted between January and March 2015. We particularly find evidence for ongoing variability when comparing the high-precision GTC data obtained in 2015 with our 2017 data (Figure \ref{fig:gtc_data}). We also find evidence of transit-like variability outside of the predicted transit windows that were determined based on the ephemeris from \citet{Sanchis2015}, but the photometric precision of our data limits us from performing a detailed investigation. Given that our highest quality data set from the GTC produced a transit that occurred at the expected time and has a duration consistent with \citet{Sanchis2015}, it is somewhat surprising to see transit-like features occurring outside of the predicted windows or lasting longer than expected. Still, our photometric monitoring campaign suggests that this disintegrating planet may potentially be outgassing at sporadic intervals since some of the transits of K2-22b that we robustly detect occur either earlier or later than predicted (Figure \ref{fig:lc_one}). 

At the epochs where we have multi-wavelength coverage during the transit windows, we find no significant color dependence of the transit light curve shape (Figure \ref{fig:gtc_color}). This lack of a color dependence is largely consistent with \citet{Sanchis2015}. However, the one color-dependent transit observed for K2-22b in 2015 is reminiscent of studies of KIC 12557548b, where \citet{bochinski2015} find evidence of deeper transits at shorter wavelengths, and of the famously dimming star KIC 8462852, where recently \citet{bodman2018arXiv180608842B}, \citet{boyajian2018}, and \citet{deeg2018} observed clear chromaticity during a dimming event. While the dimmings that occur around KIC 8462852 are more than likely due to a different phenomenon \citep[e.g., exocomets;][]{bodman2016, boyajian2016} than a disintegrating planet, that dimming was also observed to be deeper at blue than at red wavelengths \citep{bodman2018arXiv180608842B,boyajian2018,deeg2018}. \citet{bodman2018arXiv180608842B} and \citet{boyajian2018} specifically find that the chromaticity is consistent with occultation of the star by optically thin dust of the order $<<$1 $\mu$m in size on average. \citet{deeg2018} also attribute the color dependence to absorption by dust with particles sizes of $\sim$0.0015$-$0.15 $\mu$m, and they support the argument that occultation by dusty material is what causes the dimming events in that star. To strengthen the evidence for transient occurrences of dust tails and potentially even determine whether the constituents of its dust tail are similar in size to those found in KIC 12557548b's dust tail or perhaps those seen around KIC 8462852, continued monitoring of K2-22b at multiple wavelengths is warranted. Since the only color-dependent transit detected so far was also observed during one of the deepest transits of K2-22b seen to date \citep[third panel of Figure \ref{fig:gtc_data} in this paper and Figure 12 in][]{Sanchis2015}, it would be interesting to determine with future observations whether extra material being blown off the planet is what causes both a deep and color-dependent transit.

A comprehensive overview and comparison of the three known disintegrating planets around main-sequence stars was provided in \citet{vanlieshout2017}, and the data presented here further supports the fact that all disintegrating planets seem to display variable transit depths on all timescales observed, from transit to transit to several years. A continuous observing campaign by a facility capable of high-precision photometry, such as NASA's Spitzer Space Telescope, may help elucidate both the frequency at which the K2-22 system varies outside the primary transit window and how quickly the transit depth itself varies. However, Spitzer is currently expected to only operate through the end of November 2019. NASA's Transiting Exoplanet Survey Satellite \citep[TESS;][]{ricker2015,huang2018} successfully began science operations in July 2018 and could also nominally collect additional long-baseline photometry of this system. One caveat is that the orientation of the TESS field of view on the sky during its 2-year prime mission is such that it covers the ecliptic poles in full to overlap with the continuous viewing zone of NASA's James Webb Space Telescope (JWST) and largely avoids the ecliptic plane. K2-22, being in the ecliptic plane, is therefore not within the visibility window of the prime TESS mission. An extended mission for TESS in the 2020s with strategic pointings could provide the opportunity to acquire additional long-baseline time-series observations of this unique system. In the meantime, ESA's CHaracterising ExOPlanet Satellite \citep[CHEOPS;][]{broeg2013} is also expected to launch by the end of 2018 and will perform photometric follow-up of known exoplanets, potentially including K2-22. 

\citet{bodman2018arXiv180807043B} recently highlighted the unique opportunity to study the composition of K2-22b via JWST transmission spectroscopy of the planet's tail. Given the intrinsic rarity of disintegrating rocky planets like K2-22b and that the launch of JWST is in the near future (2021), now is the optimal time to prepare for such follow-up observations via additional characterization efforts with currently operating facilities.

\acknowledgments
This work has made use of observations from the LCOGT network. 
Based in part on observations made with the Gran Telescopio Canarias (GTC), installed in the Spanish Observatorio del Roque de los Muchachos of the Instituto de Astrof\'{i}sica de Canarias, in the island of La Palma.
This work has made use of data obtained at the Thai National Observatory on Doi Inthanon, operated by NARIT. This research was supported by the grant RTA5980003 from the Thailand Research Fund. 
Based in part on observations at Kitt Peak National Observatory, National Optical Astronomy Observatory (NOAO Prop. ID: 2016B-0196; PI: K. D. Col\'on), which is operated by the Association of Universities for Research in Astronomy (AURA) under a cooperative agreement with the National Science Foundation. 
Some data presented herein were obtained at the WIYN Observatory from telescope time allocated to NN-EXPLORE through the scientific partnership of the National Aeronautics and Space Administration, the National Science Foundation, and the National Optical Astronomy Observatory.
KH acknowledges support from STFC grant ST/M001296/1. 
\textsc{iraf} is distributed by the National Optical Astronomy Observatory, which is operated by the Association of Universities for Research in Astronomy (AURA) under a cooperative agreement with the National Science Foundation. 

\facilities{LCOGT, WIYN (WHIRC)}
\software{\textsc{AstroImageJ} \citep{collins2017}, Astrometry.net \citep{lang2010}, Astropy \citep{astropy2013}, \textsc{idl} \citep{landsman1993}, \textsc{Sextractor} \citep{bertin1996}, \textsc{fitsh} \citep{Pal2012}, \textsc{iraf} \citep{tody1986,tody1993}}

\bibliography{biblio}

\begin{centering}
\begin{longrotatetable}
\begin{deluxetable*}{ccccccccccc}
\tabletypesize{\footnotesize}
\tablecaption{Summary and Results of Photometric Observations of K2-22 \label{table:observations}}

\tablehead{
\colhead{Start Date} & \colhead{Transit Epoch} & \colhead{Facility} & \colhead{Aperture} & \colhead{Filter} & \colhead{Phase} & \colhead{$N_{points}$} & \colhead{Cadence} & \colhead{$\sigma$$_{median}$} & \colhead{$R_{p}/R_{*}$} & \colhead{$R_{p}/R_{*}$}\\
\colhead{(UT)} &  &  & \colhead{(m)}  & & \colhead{Coverage} & & \colhead{(min)} & \colhead{(\%)} & \colhead{(measured)} & \colhead{(corrected)}
}
\startdata
    2016-12-13 & 2427 & WIYN & 3.5 & $Ks$ & -0.144 $-$ +0.140 & 232 & 0.67 & 0.973 & 0.1142$\pm$0.0342 & $-$ \\
    2017-03-01 & 2633 & LCO/SAAO & 1.0 & $i$ & -0.154 $-$ -0.0135 & 22 & 3.67 & 0.225 &  \\
    2017-03-03 & 2638 & LCO/SAAO & 1.0 & $i$ & -0.156 $-$ +0.180 & 51 & 3.68 & 0.214 & $<$0.0426 & $<$0.0438 \\
    2017-03-04 & 2641 & LCO/SAAO & 1.0 & $i$ & -0.156 $-$ +0.186 & 52 & 3.69 & 0.231 & $<$0.0495 & $<$0.0509 \\
    2017-03-06 & 2644 & LCO/SAAO & 1.0 & $i$ & -0.155 $-$ +0.180 & 44 & 3.66 & 0.276 & $<$0.0434 & $<$0.0448 \\
    2017-03-06 & 2645 & LCO/SSO & 1.0 & $i$ & -0.156 $-$ +0.112 & 41 & 3.67 & 0.406 & $<$0.0552 & $<$0.0568 \\
    2017-03-06 & 2646 & LCO/SAAO & 1.0 & $i$ & -0.156 $-$ +0.186 & 52 & 3.67 & 0.270 & $<$0.0366 & $<$0.0376 \\
    2017-03-08 & 2650 & FLWO & 1.2 & $i$ & -0.827 $-$ -0.00465 & 198 & 2.30 & 0.192  \\
    2017-03-08 & 2650 & ULMT & 0.6 & $Clear$ & -0.587 $-$ +0.0261 & 94 & 3.54 & 0.298  \\
    2017-03-08 & 2651 & LCO/SAAO & 1.0 & $i$ & -0.156 $-$ +0.186 & 52 & 3.69 & 0.269 & $<$0.0364 & $<$0.0375  \\
    2017-03-09 & 2652 & FLWO & 1.2 & $i$ & -0.211 $-$ +0.612 & 198 & 2.30 & 0.225 & 0.0762$\pm$0.0139 & 0.0784 \\
    2017-03-09 & 2652 & ULMT & 0.6  & $Clear$ & -0.189 $-$ +0.695 & 135 & 3.54 & 0.425 & $<$0.0588 & $<$0.0612 \\
    2017-03-09 & 2654 & LCO/SAAO & 1.0 & $i$ & -0.373 $-$ -0.138 & 36 & 3.69 & 0.414 \\
    2017-03-13 & 2663 & Swope & 1.0 & $i$ & -0.708 $-$ -0.191 & 143 & 2.00 & 1.75 \\
    2017-03-14 & 2667 & LCO/SAAO & 1.0 & $i$ & -0.156 $-$ +0.0999 & 30 & 3.69 & 0.339 & $<$0.0849 & $<$0.0875  \\
    2017-03-15 & 2668 & ULMT & 0.6 & $Clear$ & -0.551 $-$ +0.430 & 146 & 3.54 & 0.491 & $<$0.1287 & $<$0.1340 \\
	2017-03-15 & 2669 & TRT-TNO & 0.5 & $R$ & -0.307 $-$ +0.174 & 250 & 1.04 & 1.13 & 0.0929$\pm$0.0184 & 0.0943 \\
    2017-03-15 & 2669 & TRT-GAO & 0.7 & $I$ & -0.304 $-$ +0.169 & 250 & 1.04 & 1.06 & 0.1035$\pm$0.0147 & 0.1084 \\
    2017-03-15 & 2669 & AAT & 3.9 & $Ks$ & -0.136 $-$ -0.0637 & 160 & 0.19 & 0.473 \\
    2017-03-15 & 2669 & LCO/SAAO & 1.0 & $i$ & +0.144 $-$ +0.305 & 24 & 3.67 & 1.28 \\
    2017-03-16 & 2671 & LCO/McDonald & 1.0 & $g$ & -0.975 $-$ -0.0629 & 104 & 3.70 & 1.44 \\
    2017-03-16 & 2671 & AAT & 3.9 & $Ks$ & -0.186 $-$ +0.624 & 3022 & 0.10 & 1.56 & 0.0572$\pm$0.0122 & 0.0615 \\
	2017-03-16 & 2672 & LCO/SAAO & 1.0 & $g$ & -0.238 $-$ +0.406 & 97 & 3.69 & 1.10 & $<$0.1001 & $<$0.1026   \\
    2017-03-16 & 2672 & TRT-GAO & 0.7 & $I$ & -0.203 $-$ +0.0680 & 140 & 1.04 & 0.880 & $<$0.09776 & $<$0.1025 \\
	2017-03-17 & 2673 & LCO/CTIO & 1.0 & $g$ & -0.372 $-$ +0.296 & 101 & 3.67 & 1.22 & 0.1362$\pm$0.0422 & 0.1396  \\
	2017-03-17 & 2673 & LCO/McDonald & 1.0 & $g$ & -0.361 $-$ +0.401 & 114 & 3.70 & 1.08 & 0.1031$\pm$0.0188 & 0.1056 \\
	2017-03-17 & 2674 & TRT-GAO & 0.7 & $I$ & -0.167 $-$ +0.123 & 150 & 1.04 & 0.647 & 0.0700$\pm$0.0120 & 0.0733  \\
	2017-03-17 & 2675 & LCO/SAAO & 1.0 & $g$ & -0.373 $-$ +0.238 & 92 & 3.69 & 1.26 & 0.1243$\pm$0.0576 & 0.1274  \\
	2017-03-18 & 2675 & LCO/CTIO & 1.0 & $g$ & +0.0391 $-$ +0.407 & 56 & 3.67 & 0.801 \\
	2017-03-18 & 2676 & LCO/McDonald & 1.0 & $g$ & -0.372 $-$ +0.168 & 80 & 3.70 & 1.05 & $<$0.0952 & $<$0.0975 \\
	2017-04-11 & 2741 & Swope & 1.0 & $i$ & -0.477 $-$ +0.233 & 145 & 2.66 & 0.473 & 0.0619$\pm$0.0224 & 0.0637 \\
	2017-04-13 & 2744 & LCO/CTIO & 1.0 & $i$ & -0.376 $-$ -0.0480 & 50 & 3.67 & 0.531 \\
	2017-04-13 & 2745 & LCO/SSO & 1.0 & $i$ & -0.375 $-$ -0.0385 & 51 & 3.69 & 0.388 \\
	2017-04-13 & 2745 & TNT-TNO & 2.4 & $i$ & -0.321 $-$ +0.197 & 221 & 0.72 & 0.345 & 0.0779$\pm$0.0168 & 0.0802 \\
	2017-04-15 & 2749 & LCO/CTIO & 1.0 & $i$ & -0.376 $-$ +0.185 & 85 & 3.67 & 0.337 & $<$0.0676 & $<$0.0695 \\
    2017-04-16 & 2752 & FLWO & 1.2 & $i$ & -0.356 $-$ -0.0368 & 78 & 2.27 & 0.190 \\
	2017-04-20 & 2762 & LCO/CTIO & 1.0 & $i$ & -0.351 $-$ -0.0504 & 46 & 3.67 & 0.284 \\
    2017-04-21 & 2765 & FLWO & 1.2 & $i$ & -0.451 $-$ -0.0278 & 103 & 2.28 & 0.173 \\
	2017-05-17 & 2835 & GTC & 10.4 & 510-1000 nm & -0.229 $-$ +0.128 & 44 & 4.56 & 0.044 & 0.0536$\pm$0.0057 & 0.0558 \\
    2017-05-22 & 2846 & FLWO & 1.2 & $i$ & -0.0780 $-$ +0.267 & 84 & 2.28 & 0.166 & $<$0.0473 & $<$0.0487 \\
    2017-05-22 & 2846 & ULMT & 0.6 & $Clear$ & -0.0665 $-$ +0.331 & 61 & 3.54 & 0.266 & $<$0.0618 & $<$0.0644  \\
    2017-05-25 & 2854 & ULMT & 0.6 & $Clear$ & -0.141 $-$ +0.162 & 43 & 3.54 & 0.292 & $<$0.0507 & $<$0.0528 \\
	2017-05-27 & 2860 & LCO/SSO & 1.0 & $i$ & -0.108 $-$ +0.100 & 32 & 3.69 & 0.373 & $<$0.0769 & $<$0.0791 \\
	2017-05-29 & 2865 & LCO/SSO & 1.0 & $i$ & -0.109 $-$ +0.0985 & 32 & 3.67 & 0.327 & $<$0.1793 & $<$0.1846  \\
	2017-05-29 & 2866 & LCO/SAAO & 1.0 & $i$ & -0.110 $-$ +0.0983 & 32 & 3.69 & 0.307 & $<$0.0526 & $<$0.0542  \\
\enddata
\tablecomments{The transit epoch is relative to the transit ephemeris given in \citet{Sanchis2015}. The cadence is defined as the median time between data points.}
\end{deluxetable*}
\end{longrotatetable}
\end{centering}

\end{document}